\documentclass{article}

\title{A closed form exact  formulation of the spectral representation of a second-order symmetric tensor and of its derivatives}

\author{Andrea Panteghini\\
DICATAM, University of Brescia\\ 
Via Branze 43, Brescia, Italy\\
email: andrea.panteghini@unibs.it}

\usepackage{graphicx}%
\usepackage{amsmath,amssymb,amsfonts}%

\newcommand {\bvec}[1]{\ds \mbox{\boldmath{$#1$}}}
\newcommand {\tens}[1]{\ds \mbox{\boldmath{$#1$}}}
\newcommand {\qtens}[1]{\ds \mathcal{#1}}

\newcommand {\ds} [0]{\displaystyle}
\newcommand {\matr} [1] {\mathbf {#1}}
\newcommand {\adj} [1] {\ds {\mbox{adj}} \left( #1 \right) }
\newcommand {\der} [2] {\ds {\frac{\partial #1}{\partial #2}}}

\newcommand {\de} [0] {\ds {\mbox{d}}}
\newcommand {\dert} [2] {\ds {\frac{{d} #1}{{d} #2}}}

\newcommand {\tr} [1] {\ds {\mbox{tr}\left( #1 \right)}}
\newcommand {\eps}{\varepsilon}

\newcommand {\myT}{T}
\newcommand {\myTdev}{t}

\begin{document}

\maketitle


\begin{abstract}
The spectral decomposition of a symmetric, second-order tensor is widely adopted in many fields of Computational Mechanics. As an example, in elasto-plasticity under large strain and rotations, given the Cauchy deformation tensor, it is a fundamental step to compute the logarithmic strain tensor. 

Recently, this approach has been also adopted in small-strain isotropic plasticity to reconstruct the stress tensor as a function of its eigenvalues, allowing the formulation of predictor-corrector return algorithms in the invariants space. These algorithms not only reduce the number of unknowns at the constitutive level, but also allow the correct handling of stress states in which the plastic normals are undefined, thus ensuring a better convergence with respect to the standard approach. 

While the eigenvalues of a symmetric, second-order tensor can be simply computed as a function of the tensor invariants, the computation of its eigenbasis can be more difficult, especially when two or more eigenvalues are coincident.
Moreover, when a Newton-Rhapson algorithm is adopted to solve nonlinear problems in Computational Mechanics, also the tensorial derivatives of the eigenbasis, whose computation is still more complicate, are required to assemble the tangent matrix.

A simple and comprehensive method is presented, which can be adopted to compute a closed form representation of a second-order tensor, as well as their derivatives with respect to the tensor itself, allowing a simpler implementation of spectral decomposition of a tensor in Computational Mechanics applications.
\end{abstract}

\section{Introduction}
This paper presents important developments regarding the eigenvalues and eigenvectors of a symmetric second-order tensor and the determination of the associated basis required for its  spectral representation.
The results here presented apply to  situations involving isotropic scalar-valued functions and isotropic tensor-valued functions of a symmetric second-order tensor.

For instance, the finding of this article are useful for the integration of constitutive laws of isotropic materials and in finite deformations (e.g.,  to compute the the logarithmic strain tensor from  the displacement gradient).

The numerical integration of isotropic elasto-plastic constitutive laws can be more efficiently carried out by formulating the return algorithms in terms of eigenvalues of the elastic strain tensor  (e.g. Borja et al. \cite{Borja2003} and de Souze Neto et al. \cite{bibbia}), or in the invariants elastic strain space \cite{Panteghini2018}, \cite{Panteghini2022}. Differently from the standard approach \cite{bibbia}, an invariant-based return algorithm allows the correct handling of stress states in which the plastic normals are undefined. 

These two integration algorithms require the spectral representation of the stress, as well as the determination its derivatives to assemble the stiffness matrix. Unfortunately, their determination using the approach described in the literature is very cumbersome (see e.g., De Souza Neto et al. \cite{bibbia}, Borja et al.  \cite{Borja2003}), particularly  when two or three eigenvalues  coincide.  This key aspect certainly makes these invariant-based integration algorithms, even if more and more efficient, less attractive with respect to standard return algorithms formulated in terms of full tensorial components.

About the applications in large strain theories, to avoid the complexity of the standard procedure, commercial codes (e.g. SIMULIA Abaqus \cite{Abaqus}) often employ approximate formulations to numerically integrate the logarithmic strain in finite deformation analyses. Some Authors suggest, for specific isotropic functions,  to resort to their numerical approximation based on series expansion (e.g. Ortiz et at. \cite{Ortiz2001}, de Souza Neto \cite{deSouzaNeto2001}, Hudobivnik et al. \cite{HUDOBIVNIK2016}). 
However, it should be noted that these series-based procedures, even if simpler and numerically efficient, can be hardly adopted when the isotropic functions are not known explicitly (i.e., for instance, in the case of the integration of the isotropic elastoplastic materials described above).

The writer has later discovered that Odgen \cite{Odgen1984} incidentally describes, in an exercise contained in his book, a very important result, which to the best of his knowledge, seems to have been missed by the vast majority of the research community. He suggests a very simple method for retrieving a closed-form expression for the basis of the spectral decomposition of a second-order tensor which does not require the computation of  the originating eigenvectors.
This result has later been reported also by Miehe \cite{Miehe1993}, who however states that " \emph{the formulation above is restricted to the case of distinct eigenvalues of the tensor}". Moreover the same Author \cite{Miehe1998} points out that such an approach requires the inversion of the second-order tensor, which severely restricts the applicability of the method. De Souza Neto et al. \cite{bibbia} describe a very cumbersome method to evaluate both the basis and their spin. They also state that "\emph{...a methodology similar to that adopted here was introduced by Miehe (1993, 1998a), where a particularly compact representation for the function derivative is used. However, the compact representation allows only the computation of the derivative at invertible arguments and cannot be used...}".

In this paper it is mathematically shown that indeed the basis required for the spectral representation of a symmetric second-order tensor can be derived without the computationally expensive evaluation of the associated eigenvectors. It is also shown that this can also be directly derived from the secular (or characteristic) equation of the tensor, without any assumptions about the invertibility of the second-order tensor. Most importantly it is clarified  how the result can be particularized to the case of two and three coinciding eigenvalues, hence removing the strong limitation of the approach described by Miehe \cite{Miehe1993}, \cite{Miehe1998} which \textit{de facto} prevents the application of this extremely useful result. This paper also provides the tensor derivatives of the basis, i.e. its \textit{spin}. 
Moreover, it is presented a simple and generic approach to compute the spectral representation of isotropic tensor-valued functions, as well as their derivatives  with respect to the tensor variable itself. The proposed procedures can be practically adopted in computational mechanics since all limitations of the procedures available in the literature have been removed (the approach of De Souza Neto et al. \cite{bibbia} does not have such limitations but is laborious to implement).  Finally two applications are presented for isotropic elasto-plasticity and for the  evaluation of the logarithmic strain tensor in finite deformations.

\section{Eigenvalues, eigenvectors and spectral representation of a symmetric, second-order tensor $\tens{\myT}$}
Given the symmetric, second-order tensor $\tens{\myT}$, its (ordered) eigenvalues $\lambda_i$ and their corresponding eigenvectors $\bvec{n}_i$ are obtained
by solving  the eigenvalues-eigenvectors problem \cite{malvern1969}:
\begin{equation}
\left(\tens \myT -\lambda \tens I \right)\bvec{n} = \bvec 0\; \mbox{ under the condition } \;\bvec{n}^T\bvec{n}=1
\label{eq:eigen_problem}
\end{equation}
being $\tens{I}$  the second-order identity tensor. 
The principal components $\lambda_i$ can be obtained by solving the third-order scalar equation in $\lambda$, namely the \emph{secular equation}: 
\begin{equation}
\lambda^3 -I_1 \lambda^2 + I_2 \lambda - I_3 = 0  
\label{eq:secular}
\end{equation}
The coefficients
\begin{equation}
I_1 = \tr{\tens{\myT}}
\label{eq:I1}
\end{equation}
\begin{equation}
I_2 = \frac{1}{2}\left(I_1^2 - \tens{\myT}:\tens{\myT}^T\right)
\label{eq:I2}
\end{equation}
\begin{equation}
I_3 = \det \left(\tens{\myT} \right)
\label{eq:I3}
\end{equation}
are the \emph{invariants} of $\tens{\myT}$, since their values do not depend on the reference system in which $\tens{\myT}$ is expressed. The three \emph{ordered} solutions  of  Eq. \eqref{eq:secular} are the eigenvalues of the problem described in Eq. \eqref{eq:eigen_problem}. As explained in \cite{malvern1969}, they can be computed in closed form as:
\begin{equation}
\begin{gathered}
\lambda_{\rm{I}}=\frac{I_1}{3}+\frac{2}{\sqrt{3}} \sqrt{J_2} \sin{\left(\theta+ \frac{2}{3} \pi\right)}\\
\lambda_{\rm{II}}=\frac{I_1}{3}+\frac{2}{\sqrt{3}} \sqrt{J_2}  \sin{\left(\theta\right)}\\
\lambda_{\rm{III}}=\frac{I_1}{3}+\frac{2}{\sqrt{3}} \sqrt{J_2}  \sin{\left(\theta- \frac{2}{3} \pi\right)}
\end{gathered}
\label{eq:lambda}
\end{equation}
where
\begin{equation}
J_2 = \frac{1}{2} \tens{\myTdev}: \tens{\myTdev}
\label{eq:defj2}
\end{equation}
\begin{equation}
J_3 = \det \left(\tens{\myTdev} \right)
\nonumber
\end{equation}
are the invariants of the second-order, deviatoric symmetric tensor $\tens{\myTdev}=\tens \myT - \frac{I_1}{3} \tens I$, and the Lode's angle $\theta$ is defined as
\begin{equation}
\theta = \frac{1}{3} \arcsin\left( - \frac{\sqrt{27}}{2}  \frac{J_3}{\sqrt{J_2^3}}\right)
\label{eq:deftheta}
\end{equation}
where $-\pi/6 \leq \theta \leq \pi/6$.

It is well known that the second-order symmetric tensor $\tens{T}$ can be expressed as a function of its eigenvalues $\lambda_i$ and the corresponding eigenvectors $\bvec n_i$ by resorting to the spectral theorem\footnote{
Let consider that, unless otherwise specified, it is always intended
\begin{equation*}
\ds \sum  f_i = \ds \sum_{i=\rm{I},\rm{II},\rm{III}} f_i
\end{equation*}
}:

\begin{equation}
    \tens T = \sum \lambda_i \bvec{n}_i \otimes \bvec{n}_i = \lambda_i \bvec N_i
    \label{eq:spectral_intr}
\end{equation}
where $\tens{N}_i$ is the eigenbasis of $\bvec T$ related to $\lambda_i$.

\section{Closed-form expression for the eigenbasis of $\tens T$}

We will consider three cases, as a function of the multiplicity of the eigenvalues $\lambda_i$:
\begin{enumerate}
    \item $\lambda_{\rm{I}}> \lambda_{\rm{II}} >\lambda_{\rm{III}}$
    \item  $\lambda_{\rm{I}}> \lambda_{\rm{II}}=\lambda_{\rm{III}}$ or 
    $\lambda_{\rm{I}}= \lambda_{\rm{II}}>\lambda_{\rm{III}}$
    \item $\lambda_{\rm{I}}= \lambda_{\rm{II}}=\lambda_{\rm{III}}$
\end{enumerate}

Let observe that the number of non coincident eigenvalues, i.e., the the eigenvalues multiplicity can be simply determined from the invariants of $\tens T$. Hence,
case (i) occurs when $J_2\neq0$ and $ \theta \ne \pm \pi/6$,  the case (ii) implies $J_2\neq0$ and $\theta=\pm \pi/6$, and finally the case (iii) requires that $J_2=0$ (while $\theta$ is undefined).
\subsubsection*{A general property of the eigenbasis $\tens N_i$.}
We will initially prove that it results:
\begin{equation}
\sum \tens{N}_i = \tens{I}
\label{eq:sum_eigenbasis}
\end{equation}
Let consider that the $i-$th eigenvalue and eigenvector of $\tens{\myT}$ will satisfy Eq. \eqref{eq:eigen_problem}, i.e.
\begin{equation}
   \tens \myT  \bvec{n}_{i}= \lambda_{\rm{i}} \bvec{n}_{\rm{i}} 
   \label{eq:eigen_sol}
 \end{equation}
Since $\bvec{n}_i$ is a unit vector, it results
\begin{equation}
\bvec{n}_i^T \tens \myT \bvec{n}_i = 
\tens \myT : \left( \bvec{n}_i \otimes \bvec{n}_i\right)= \lambda_i \left( \bvec{n}_i^T \bvec{n}_i\right)= \lambda_i
\nonumber
\end{equation}
one can compute the first invariant $I_1$ in the principal coordinate system as
\begin{equation}
I_1=\tr{\tens \myT}=\tens \myT:\tens{I}=\sum \lambda_i =  \tens \myT : \sum \tens{N}_i
\nonumber
\end{equation}
 From this equation it must result
\begin{equation}
\tens \myT:\tens{I} = \tens \myT : \sum \tens{N}_i
\nonumber
\end{equation}
This conditions yields
\begin{equation}
\sum \tens{N}_i = \tens{I}
\nonumber
\end{equation}

\subsubsection*{Case (i):  $\lambda_{\rm{I}}> \lambda_{\rm{II}} >\lambda_{\rm{III}}$.}
One will prove that the spectral theorem
    \begin{equation}
\tens{\myT} =\sum \lambda_i \left(\bvec{n}_i \otimes \bvec{n}_i \right) = \sum \lambda_i \tens{N}_i
\label{eq:spectral}
\end{equation}
can be written as
\begin{equation}
\tens{\myT}=   \sum \lambda_i \dert{\lambda_i}{\tens{\myT}}
\nonumber
\end{equation}
i.e., we will prove that it simply results\footnote{It should be noted that, to the best of the Author's knowledge, this result appears for the first time, without any demonstration or explanation in Ogden's book \cite{Odgen1984}.
It has been used by Mihe \cite{Miehe1993}, \cite{Miehe1998}, but, as explained in the Introduction, due to the limitations of his approach, it seems it is not commonly adopted in Computational Mechanics.
}:
\begin{equation}
\tens{N}_i=    \dert{\lambda_i}{\tens{\myT}}
\nonumber
\end{equation}
By considering the symmetry of $\tens{T}$, the derivatives of the invariants $I_1$, $I_2$ and $I_3$, defined by Eq. \eqref{eq:I1}, \eqref{eq:I2} and \eqref{eq:I3} with respect to $\tens \myT$ are:
\begin{equation}
\dert{I_1}{\tens{\myT}}=\tens{I}
\label{eq:derI1std}
\end{equation}
\begin{equation}
\dert{I_2}{\tens{\myT}}=I_1 \tens{I}-\tens{\myT}
\label{eq:derI2std}
\end{equation}
\begin{equation}
\dert{I_3}{\tens{\myT}}=I_3 \tens{\myT}^{-1} =\adj{\tens{\myT}}
\label{eq:derI3std}
\end{equation}
where $\adj{\tens{\myT}}$ denotes the adjugate matrix of $\tens{\myT}$.
By substituting the property \eqref{eq:sum_eigenbasis} and the spectral theorem \eqref{eq:spectral} into Eq. \eqref{eq:derI1std} and \eqref{eq:derI2std} respectively, one obtains:
\begin{equation}
\dert{I_1}{\tens{\myT}}=\sum \tens{N}_i
\label{eq:derI1mod}
\end{equation}
\begin{equation}
\dert{I_2}{\tens{\myT}}=I_1 \tens{I}-\sum \lambda_i \tens{N}_i
\label{eq:derI2mod}
\end{equation}
Finally, by resorting to the spectral theorem \eqref{eq:spectral}, one can write \eqref{eq:derI3std} as
\footnote{
Let observe that, by multiplying Eq. \eqref{eq:eigen_problem} by $\adj{\tens{\myT}} = I_3 \tens{\myT}^{-1}$ one obtains
\begin{equation}
I_3 \tens{\myT}^{-1} \tens{\myT} \bvec{n}= \lambda I_3 \tens{\myT}^{-1} \bvec{n} 
\nonumber
\end{equation}
which gives
\begin{equation}
\adj{\tens{\myT}} \bvec{n} = \frac{I_3}{\lambda} \bvec{n}
\end{equation}
Hence, the eigenvectors $\bvec{n}$ of $\adj{\tens{\myT}}$ and $\tens{\myT}$ are coincident, whilst the $i$-th eigenvalue $\mu_i$ of $\adj{\tens{\myT}}$ associated to $\bvec n_{i}$ can be computed from $\lambda_i$ as:
\begin{equation}
\mu_i= \frac{I_3}{\lambda_i}= \left(\lambda_{j} \lambda_{k} \right)_{i \ne j \ne k}
\end{equation}
The spectral representation of $\adj{\tens{\myT}}$ is then:
\begin{equation}
    \adj{\tens{\myT}}= \sum \left( \lambda_{j} \lambda_{k}  \tens{N}_i \right)_{i \ne j \ne k}
\end{equation}
}

\begin{equation}
\dert{I_3}{\tens{\myT}}= \sum \left(\lambda_{j} \lambda_{k}  \tens{N}_i \right)_{i \ne j \ne k}
\label{eq:derI3mod}
\end{equation}

Let consider now that the value of $I_1$, $I_2$ and $I_3$ are independent with respect to the reference systems, hence one can compute them also in terms of principal components. It result:
\begin{equation}
I_1= \lambda_{\rm{I}}+\lambda_{\rm{II}}+\lambda_{\rm{III}}
\nonumber
\end{equation}
\begin{equation}
I_2= \lambda_{\rm{I}} \lambda_{\rm{II}}+ \lambda_{\rm{I}} \lambda_{\rm{III}} +\lambda_{\rm{II}} \lambda_{\rm{III}} 
\nonumber
\end{equation}
\begin{equation}
I_3 = \lambda_{\rm{I}} \lambda_{\rm{II}} \lambda_{\rm{III}}
\nonumber
\end{equation}
The derivatives of the invariants  $I_1$, $I_2$ and $I_3$ can also be computed by differentiating these last three expressions, observing that $\lambda_i=\lambda_i(\tens \myT)$. It results:
\begin{equation}
\dert{I_1}{\tens{\myT}}=\dert{\lambda_{\rm{I}}}{\tens{\myT}}+\dert{\lambda_{\rm{II}}}{\tens{\myT}}+\dert{\lambda_{\rm{III}}}{\tens{\myT}} =\sum \dert{\lambda_{i}}{\tens{\myT}}
\label{eq:derI1inv}
\end{equation}
\begin{equation}
\dert{I_2}{\tens{\myT}}=I_1\sum \dert{\lambda_{\rm{i}}} {\tens{\myT}}-\sum \lambda_{i} \dert{\lambda_{\rm{i}}} {\tens{\myT}}= I_1 \tens{I} -\sum \lambda_{i} \dert{\lambda_{\rm{i}}} {\tens{\myT}}
\label{eq:derI2inv}
\end{equation}
\begin{equation}
\dert{I_3}{\tens{\myT}} = \lambda_{\rm{II}}
\lambda_{\rm{III}} \dert{\lambda_{\rm{I}}}{\tens{\myT}}
+
\lambda_{\rm{I}}
\lambda_{\rm{III}} \dert{\lambda_{\rm{II}}}{\tens{\myT}}
+
\lambda_{\rm{I}}
\lambda_{\rm{II}} \dert{\lambda_{\rm{III}}}{\tens{\myT}}
=\sum \left(\lambda_{j} \lambda_{k} \dert{\lambda_{i}}{\tens{\myT}}  \right)_{i \ne j \ne k}
\label{eq:derI3inv}
\end{equation}

One can now compute the eigenbasis $\tens{N}_{i}$ as a function of the derivatives of the eigenvalues $\lambda_i$ with respect to $\tens{\myT}$ by solving the linear system of equations obtained by equating Eq. \eqref{eq:derI1mod}, \eqref{eq:derI2mod} and \eqref{eq:derI3mod} with Eq. \eqref{eq:derI1inv}, \eqref{eq:derI2inv} and \eqref{eq:derI3inv} respectively.
One obtains
\begin{equation}
\left\{
\begin{aligned}
&  \sum \tens{N}_i  = \sum \dert{\lambda_{\rm{i}}} {\tens{\myT}}\\
&  \sum \lambda_i \tens{N}_i  = \sum \lambda_i \dert{\lambda_{\rm{i}}} {\tens{\myT}}\\
& \sum \left(\lambda_{j} \lambda_{k}  \tens{N}_i \right)_{i \ne j \ne k}=\sum \left(\lambda_{j} \lambda_{k} \dert{\lambda_{i}}{\tens{\myT}}  \right)_{i \ne j \ne k}
\end{aligned}
\right.
\label{eq:sist1}
\end{equation}
which, under the assumption $\lambda_{\rm{I}}> \lambda_{\rm{II}} >\lambda_{\rm{III}}$\footnote{
Let observe that the determinant of the matrix of the system \eqref{eq:sist1} reads:
\begin{equation}
\begin{aligned}
\det \begin{bmatrix}
    1 & 1 & 1\\
    \lambda_{\rm{I}} & \lambda_{\rm{II}} & \lambda_{\rm{III}}\\
    \lambda_{\rm{II}}\lambda_{\rm{III}} & \lambda_{\rm{I}}\lambda_{\rm{III}} & \lambda_{\rm{I}} \lambda_{\rm{II}}
\end{bmatrix}
= - \left(\lambda_{\rm{I}}-\lambda_{\rm{II}}\right)\left(\lambda_{\rm{I}}-\lambda_{\rm{III}}\right)\left(\lambda_{\rm{II}}-\lambda_{\rm{III}}\right)
\nonumber
\end{aligned}
\end{equation}
It is always nonzero if $\lambda_{\rm{I}}> \lambda_{\rm{II}} >\lambda_{\rm{III}}$.
} simply gives
\begin{equation}
\tens{N}_{i} =  \dert{\lambda_i}{\tens{\myT}}
\nonumber
\end{equation}
so that the spectral theorem \eqref{eq:spectral} can be re-written as:
\begin{equation}
\tens{\myT}= \sum \lambda_i \left(\bvec{n}_i \otimes \bvec{n}_i \right) =  \sum \lambda_i \dert{\lambda_i}{\tens{\myT}}
\nonumber
\end{equation}

\subsubsection*{Case (ii): $\lambda_{\rm{I}}> \lambda_{\rm{II}}=\lambda_{\rm{III}}$ or 
    $\lambda_{\rm{I}}= \lambda_{\rm{II}}>\lambda_{\rm{III}}$.}

If one or more eigenvalues are coincident of $\tens{\myT}$, then the linear system \eqref{eq:sist1} will not admit a unique solution.   
Let  $\hat \lambda$ be the non-repeated eigenvalue of $\tens{\myT}$ and $\hat{\tens{N}}$ the correspondent eigenbasis.
The first invariant $I_1$ is equal to:
\begin{equation}
I_1 = \hat \lambda + 2 \lambda_{\rm{II}}
\nonumber
\end{equation}
so that, it results:
\begin{equation}
 \lambda_{\rm{II}}= \frac{1}{2} \left(I_1 - \hat \lambda\right)
 \nonumber
\end{equation}
Eq. \eqref{eq:sum_eigenbasis} can be rewritten as:
\begin{equation}
\hat{\tens{N}}  + 2  \tens{N}_{\rm{II}} = \tens{I}
\nonumber
\end{equation}
hence, it results:
\begin{equation}
  \tens{N}_{\rm{II}} = \frac{1}{2}\left(\tens{I} -\hat{\tens{N}} \right)
\label{eq:eigenbasis_2coinc}
\end{equation}
The spectral theorem can be rewritten as:
\begin{equation}
\begin{aligned}
\tens{\myT}=\hat \lambda\hat{\tens{N}} + \frac{1}{2} \left(I_1-\hat \lambda \right) \left(\tens{I} -\hat{\tens{N}} \right)
=
\frac{3}{2} \left(\hat \lambda - \frac{I_1}{3} \right) \hat{\tens{N}}+ \frac{1}{2} \left( I_1- \hat \lambda \right) \tens{I}
\end{aligned}
\label{eq:spectral2}
\end{equation}
Eq. \eqref{eq:spectral2} can be further simplified by computing the deviatoric part $\hat{l}$ of $\hat \lambda$ as $\hat{l}=\hat \lambda-I_1/3$. One obtains
\begin{equation}
\tens{\myT}= \frac{I_1}{3} \tens{I}+ \frac{3}{2}\hat{l} \left( 
\tens{N}_{j} -\frac{1}{3} \tens{I} \right)
\label{eq:spectral_dev}
\end{equation}
This last equation clearly shows that, when two eigenvalues are coincident, the deviatoric part of $\hat{\tens{N}}$, defined as $\hat{\tens{N}}^d=  \hat{\tens{N}} -\tens{I}/3$, \emph{is simply proportional} to the deviatoric part of the tensor $\tens{\myT}$, i.e.
\begin{equation}
\begin{gathered}
\hat{\tens{N}}^d = \frac{1}{\hat \lambda -\lambda_{\rm{II}}}\tens{t} = \mp \frac{1}{q} \tens{t} \;\;\mbox{ for } \;\; \theta=\pm \frac{\pi}{6}
\label{eq:spectral_invariants3}
\end{gathered}
\end{equation}
where $q=\sqrt{3 J_2}$.
This result is a consequence of the multiplicity of the \emph{deviatoric} principal components.
When two eigenvalues of $\tens{T}$ coincide, the two coincident deviatoric principal components result to be
 minus half of the (only) independent one, since their sum must vanish.
Eq. \eqref{eq:spectral2} results to be the sum of two independent terms: the volumetric and the deviatoric parts. The \emph{basis} of the volumetric part is obviously proportional to the identity tensor $\tens I$, whilst that of the deviatoric part can only be proportional to the tensor itself.
\begin{equation}
    \tens{\myT}= \frac{I_1}{3} \tens{I}+ \frac{3}{2} \hat{l} \hat{\tens{N}}^d 
    \nonumber
\end{equation}
It should be noted that, as in Case (i), it is still possible to demonstrate that 
\begin{equation}
   \hat{\tens{N}}=\dert{\hat \lambda}{\tens{\myT}}
   \nonumber
\end{equation}
To prove this result, let compute the second invariant $J_2$ of the deviatoric tensor $\tens{\myTdev}$ as a function of the principal component $\hat \lambda$:
\begin{equation}
\begin{aligned}
    J_2= \frac{\tens{\myTdev}:\tens{\myTdev}}{2}
    =\frac{\left(\hat \lambda -I_1/3 \right)^2 + 2 \left(\lambda_{\rm{II}}  -I_1/3 \right)^2 }{2}
    = \frac{3 \left(\hat \lambda -I_1/3 \right)^2 }{4}
    \end{aligned}
    \label{eq:j2basis}
\end{equation}
By differentiating this expression with respect to $\tens{\myT}$, one obtains
\begin{equation}
\dert{J_2}{\tens{\myT}}= \tens{\myTdev} = \tens{\myT} - \frac{I_1}{3} \tens{I}=\frac{3}{2} \left(\hat \lambda -\frac{I_1}{3} \right) \left(\dert{\hat \lambda}{\tens{\myT}}- \frac{1}{3}\tens{I} \right)
\label{eq:derj2}
\end{equation}
so that, solving for $\tens{\myT}$ one obtains:
\begin{equation}
\tens{\myT}= \frac{3}{2} \left(\hat \lambda - \frac{I_1}{3} \right)\dert{\hat \lambda}{\tens{\myT}}+ \frac{1}{2} \left( I_1- \hat \lambda \right) \tens{I}
\nonumber
\end{equation}
By equating this last expression with Eq. \eqref{eq:spectral2} and solving for $\hat{\tens{N}}$\footnote{This can be done under the condition $\hat \lambda \ne I_1/3$ that, observing Eq. \eqref{eq:j2basis} is equivalent to $J_{2}\ne0$}
 one obtains:
\begin{equation}
   \hat{\tens{N}} = \dert{\hat \lambda}{\tens{\myT}}
\label{eq:dernJ}
\end{equation}

\subsubsection*{Case (iii): $\lambda_{\rm{I}}= \lambda_{\rm{II}}=\lambda_{\rm{III}}$.}
Finally, let consider the case of three coincident eigenvalues $\lambda=\lambda_{\rm{I}}= \lambda_{\rm{II}}=\lambda_{\rm{III}}$. The tensor $\tens{\myT}$ is purely volumetric in any reference system. By observing that it results $l_i = 0\;\; \forall i$ and $I_1=3\lambda$, Eq. \eqref{eq:spectral_dev}  simply becomes
\begin{equation}
\tens{\myT}= \lambda \tens{I}
\label{eq:spectral3}
\end{equation}
From Eq. \eqref{eq:sum_eigenbasis} it results
\begin{equation}
\tens{N}_{\rm{I}}=\tens{N}_{\rm{II}}=\tens{N}_{\rm{III}}=\frac{1}{3}\tens{I}
\nonumber
\end{equation}

\section{Computation the eigenbasis directly from the secular equation}

Since the three eigenbasis are equal to the derivatives of its conjugate principal components with respect to the tensor $\tens{\myT}$, one can determine them by simply differentiating Eqs. \eqref{eq:lambda} with respect to $\tens{\myT}$. Using the chain rule, one obtains:
\begin{equation}
\tens{N}_{i} = \dert{\lambda_i}{\tens{\myT}} 
= \frac{1}{3} \tens{I}+ \frac{\sqrt{3}}{3} \left( \frac{\sin \beta_i}{\sqrt{J_2}} \dert{J_2}{\tens{\myT}}+2 J_2 \cos \beta_i \dert{\theta}{\tens{\myT}}
\right)
\nonumber
\end{equation}
where $\beta_{\rm{I}}= \theta+2/3 \pi$,  $\beta_{\rm{II}}= \theta$, $\beta_{\rm{III}}= \theta-2/3 \pi$, and
\footnote{
It should be noted that Eq. \eqref{eq:dertheta} requires the computation of $\tens{\myTdev}^{-1}$. An expression more suitable for the implementation is
\begin{equation}
\dert{\theta}{\tens{\myT}}=-\frac{1}{\cos 3 \theta}
\left(\frac{\sqrt{3}}{2 \sqrt{J_2^3}} \dert{J_3}{\tens{\myTdev}}+ \frac{\sqrt{3}}{6 \sqrt{J_2}} \tens{I} + \frac{\sin 3\theta}{2 J_2} \tens{\myTdev} \right)
\label{eq:dertheta_impl}
\end{equation}
where
\begin{equation}
\dert{J_3}{\tens{\myTdev}}=
\begin{bmatrix}
s_{yy} s_{zz}-s_{yz}^2 & s_{xz} s_{yz}- s_{xy} s_{zz} & s_{xy} s_{yz} - s_{xz} s_{yy} \\
 s_{xz} s_{yz}- s_{xy} s_{zz} & s_{xx} s_{zz} -s_{xz}^2 & s_{xy} s_{xz} - s_{yz} s_{xx}\\
 s_{xy} s_{yz} - s_{xz} s_{yy} & s_{xy} s_{xz} - s_{yz} s_{xx} & s_{xx} s_{yy} - s_{xy}^2
\end{bmatrix}
\nonumber
\end{equation}
that is undefined only for $J_2=0$ or $\theta=\pm \pi/6$
}
\begin{equation}
\dert{J_2}{\tens{\myT}}= \tens{\myTdev}
\nonumber
\end{equation}
\begin{equation}
\dert{\theta}{\tens{\myT}}=\frac{1}{\cos 3 \theta}
\left(\frac{\sin 3\theta}{3} \tens{\myTdev}^{-1} - \frac{\sqrt{3}}{6 \sqrt{J_2}} \tens{I} - \frac{\sin 3\theta}{2 J_2} \tens{\myTdev} \right)
\label{eq:dertheta}
\end{equation}
The computation of the spin of the eigenbasis, i.e. $\dert{\tens{N}_{i} }{\tens{\myT}}$
is even more tiring.

A more elegant and simpler approach can be obtained by working directly on the secular equation \eqref{eq:secular}. Each of the eigenvalues $\lambda_i$ will satisfy Eq. \eqref{eq:secular}, i.e.
\begin{equation}
f(\tens{\myT})= \lambda_i^3 - I_1 \lambda_i^2 + I_2 \lambda_i- I_3 =0
\nonumber
\end{equation}
hence, it must result 
\begin{equation}
\begin{aligned}
\de f(\tens{\myT})=\left[\left(3 \lambda_i^2 - 2 I_1 \lambda_i + I_2 \right) \dert{\lambda_i}{\tens{\myT}}- \tens{I} \lambda_i^2 \right. \\
\left. + \left(I_1 \tens{I}  - \tens{\myT} \right)\lambda_i + I_3 \tens{\myT}^{-1}\right]: \de \tens{\myT}=0 \;\; \forall \;\de \tens{\myT}
\end{aligned}
\nonumber
\end{equation}
This imply the condition:
\begin{equation}
\begin{aligned}
\left(3 \lambda_i^2 - 2 I_1 \lambda_i + I_2 \right) \dert{\lambda_i}{\tens{\myT}}- \tens{I} \lambda_i^2 + \left(I_1 \tens{I}  - \tens{\myT} \right)\lambda_i 
+ I_3 \tens{\myT}^{-1}=0
\end{aligned}
\nonumber
\end{equation}
 The eigenbasis $\tens{N}_i$ can be obtained by simply solving this last equation of $\dert{\lambda_i}{\tens{\myT}}$. By observing that $J_2=1/3 I_1^2-I_2$, after some simple algebraic manipulation, one obtains\footnote{
 A very compact way to write this derivative is $I_3 \tens{\myT}^{-1}$. However, it should be noted that it is not completely correct from a formal point of view, since it is  undefined when $I_3=0$. The invariant $I_3$, being defined as $\det \tens{\myT}$, is simply the adjugate matrix of $\tens \myT$, that is always defined. In simpler words, being $I_3=\det \myT$ 
 a third degree polynomial in $\myT_{ij}$, its derivative with respect to $\tens{\myT}$ is always defined. It results:
\begin{equation}
\begin{aligned}
\dert{I_3}{\tens{\myT}}=\adj {\tens {\myT}}
=\begin{bmatrix}
\myT_{yy} \myT_{zz}-\myT_{yz}^2 & \myT_{xz} \myT_{yz}- \myT_{xy} \myT_{zz} & \myT_{xy} \myT_{yz} - \myT_{xz} \myT_{yy} \\
 \myT_{xz} \myT_{yz}- \myT_{xy} \myT_{zz} & \myT_{xx} \myT_{zz} -\myT_{xz}^2 & \myT_{xy} \myT_{xz} - \myT_{yz} \myT_{xx}\\
 \myT_{xy} \myT_{yz} - \myT_{xz} \myT_{yy} & \myT_{xy} \myT_{xz} - \myT_{yz} \myT_{xx} & \myT_{xx} \myT_{yy} - \myT_{xy}^2
\end{bmatrix}
\end{aligned}
\nonumber
\end{equation}
Eq. \eqref{eq:eigenbasis} becomes
\begin{equation}
 \tens{N}_{i} = \dert{\lambda_i}{\tens{\myT}} = \frac{ \lambda_i \left[\left(\lambda_i-I_1  \right) \tens{I}+  \tens{\myT}\right]+\dert{I_3} {\tens{\myT}}}{J_2 \left(4 \sin^2 \beta_i -1\right)}
 \label{eq:eigenbasis_impl}
 \end{equation}
}:
\begin{equation}
\tens{N}_{i} = \dert{\lambda_i}{\tens{\myT}} = \frac{ \lambda_i \left[\left(\lambda_i-I_1  \right) \tens{I}+  \tens{\myT}\right]+I_3\tens{\myT}^{-1}}{J_2 \left(4 \sin^2 \beta_i -1\right)}
\label{eq:eigenbasis}
\end{equation}

The spin of the eigenbasis can be obtained by differentiating Eq. \eqref{eq:eigenbasis} by the tensor $\tens{\myT}$. One obtains
\begin{equation}
\begin{split}
\dert{\tens{N}_{i}}{\tens{\myT}}=
\frac{\mbox{d}^2\lambda_i}{\mbox{d}\tens{\myT} \otimes \mbox{d}\tens{\myT}}
=\frac{1}{J_2 \left(4 \sin^2 \beta_i -1\right)}
\left[ - 4 \sqrt{3 J_2} \sin \beta_i \left(\tens{N}_{i} \otimes \tens{N}_{i} \right) \right. \\ \left. 
+
\left(2 \lambda_i -I_1 \right) \left (\tens{N}_{i} \otimes \tens{I} 
+ \tens{I} \otimes \tens{N}_{i}  \right)\right.\\
\left.+   \left (\tens{N}_{i} \otimes \tens{\myT} 
+ \tens{\myT} \otimes \tens{N}_{i} \right)
+
\lambda_i \left(\qtens{I} -\tens{I}\otimes \tens{I} \right)
+ \frac{\mbox{d}^2 I_3}{\mbox{d}\tens{\myT} \otimes \mbox{d}\tens{\myT}}
\right]
 \end{split}
 \label{eq:spin}
\end{equation}
where $\qtens{I}$ is the fourth-order identity tensor and
\begin{equation}
\frac{\mbox{d}^2 I_3}{\mbox{d}\tens{\myT} \otimes \mbox{d}\tens{\myT}} = \delta_{jk} \myT_{il}+ \myT_{jk} \delta_{il}
\nonumber
\end{equation}
being $\delta_{ij}$  the Kroneker delta operator.

Let note that, even in the case of two coincident $\lambda_i$, the spin of the basis associated to the non-repeated eigenvalue $\hat \lambda$ can still be computed using Eq. \eqref{eq:spin}. It is the only spin required to compute the derivative of  Eq. \eqref{eq:spectral_dev}. 
However, by exploiting the \emph{proportionality} between the deviatoric part of the tensor and the basis itself, it can be simpler obtained by means of Eq. \eqref{eq:spectral_invariants3}.
As explained in the previous section, when all the eigenvalues coincide, the three eigenbasis $\tens{N}_i$ are simply equal to $\tens{I}/3$. Their spin is not defined, but, as explained in the next section, it is still possible to evaluate the derivative of the spectral representation of the tensor when its invariants are isotropic functions.

\section{Isotropic functions}

In many  mechanical applications it is \emph{a priori} known that two second-order, symmetric tensors $\tens S$ and $\tens T$ share the same principal directions. Under these conditions, the two tensors are called \emph{co-axial}.
These applications usually involve isotropic tensor functions, i.e., the invariants of the tensor $\tens T$ are function of the those of the tensor $\tens S$.

In these applications, once the principal components $\eta_i$ of the tensor $\tens S$ are computed as a function of those of $\tens T$, say $\lambda_{i}$ it is  finally required to compute the Cartesian components of $\tens S$.



Let $\tens{S}$ be a symmetric, second-order tensor, co-axial with $\tens{\myT}$.
Let assume that the  generic eigenvalues $\eta_{i}( \lambda_{\rm{I}}, \lambda_{\rm{II}}, \lambda_{\rm{III}})$  of $\tens{S}$ can be computed as a function of the eigenvalues $\lambda_i$ of $\tens{\myT}$. 
Since $\tens{S}$ and $\tens{\myT}$ are co-axial, they will share the same eigenbasis $ \tens{N}_i$ and it results
\begin{equation}
    \tens{S}=\sum \eta_i ( \lambda_{\rm{I}}, \lambda_{\rm{II}}, \lambda_{\rm{III}})\tens{N}_i
    \nonumber
\end{equation}
Since it results $\tens{N}_i \otimes \tens{N}_j=\tens{0}$ for $i\ne j$, the derivative of this expression with respect to the tensor $\tens{T}$ will be
\begin{equation}
    \dert{\tens{S}}{\tens{\myT}} = \sum \der{\eta_i}{\lambda_i} \tens{N}_i \otimes \tens{N}_i + \eta_i 
    \dert{\tens{N}_{i}}{\tens{\myT}}
    \label{eq:jacobian:compl}
\end{equation}
Let consider the case in which two eigenvalues $\lambda_i$ of $\tens{\myT}$ coincide.
As explained in the section above, under this condition it results that the deviatoric part of $\tens{S}$, say $\tens{s}$, results to be \emph{proportional} to the deviatoric part of $\tens{T}$, say $\tens{t}$. Hence, one can compute $\tens{S}$ as
\begin{equation}
    \tens{S}= \frac{I_{1S}}{3} \tens{I}+ \frac{q_S}{q_T} \tens{t}
    \label{eq:isotr2coinc}
\end{equation}
where $I_{1T}=\tr{\tens{T}}$ is the first invariant of $\tens{T}$, $q_T=\sqrt{\frac{3}{2} \tens{t}:\tens{t}}$, $I_{1S}=\tr{\tens{S}}=I_{1S}(I_{1T},q_T)$ and $q_S=\sqrt{\frac{3}{2} \tens{s}:\tens{s}}=q_{S}(I_{1T},q_T)$.

Let now compute $\dert{\tens{S}}{\tens{\myT}}$. Since $\tens{s}$ and $\tens{t}$ are simply proportional, it must result
\begin{equation}
\theta_S=\theta_T
\nonumber
\end{equation}
and then
\begin{equation}
    \der{\theta_S}{\theta_T}=1
    \nonumber
\end{equation}
Moreover, considering that 
 Eq. \eqref{eq:stressinvar} gives:
\begin{equation}
q_S(\theta_S)= \sqrt[3]{-\frac{27}{2} \frac{J_{3S}}{\sin (3 \theta_S)}}
\nonumber
\end{equation}
it results
\begin{equation}
\begin{aligned}
\der{q_S}{\theta_T}=\der{q}{\theta_{S}}\der{\theta_S}{\theta_T}
= \frac{3\sqrt[3]{4}}{2}
\frac{J_{3S} \cos (3 \theta_S) \sqrt[3]{\sin^2 (3 \theta_S)}}
{\sqrt[3]{J_{3S}^2} \sin^2 (3 \theta_S)}= 0 
\mbox{ for } \theta_S=\theta_T=\pm \frac{\pi}{6}
\end{aligned}
\nonumber
\end{equation}
Analogously
\begin{equation}
\der{I_{1S}}{\theta_{T}}=\der{I_{1S}}{\theta_S}\der{\theta_S}{\theta_T}=0 \;\;\; \forall \theta_T
\nonumber
\end{equation}
Hence, observing that from Eq. \eqref{eq:spectral_invariants3} it results that 
\begin{equation}
    \dert{q_T}{\tens{T}} = \frac{3}{2 q_T} \tens{t} = \mp \frac{3}{2}   \hat{\tens{N}}^d
    \;\; \mbox{ for }\;\; \theta_T=\pm \frac{\pi}{6}
\end{equation}
by differentiating Eq. \eqref{eq:isotr2coinc} with respect to $\tens{T}$ one obtains:
\begin{equation}
\begin{aligned}
    \dert{\tens{S}}{\tens{\myT}}= \frac{1}{3}\der{I_{1S}}{I_{1T}} \tens{I} \otimes \tens{I} \mp \frac{1}{2} \der{I_{1S}}{q_T} \tens{I} \otimes \hat{\tens{N}}^d
    + \frac{q_S}{q_T}\left(\mathcal{I} - \frac{1}{3}   \tens{I} \otimes \tens{I}  \right)\\
    + \frac{3}{2} \left(\der{q_S}{q_T} 
      - \frac{q_S}{q_T} \right) \hat{\tens{N}}^d \otimes \hat{\tens{N}}^d 
    \mp\der{q_S}{I_{1T}}  \hat{\tens{N}}^d \otimes \tens{I}\;\; \mbox { for }\;\; \theta_T = \pm \frac{\pi}{6}
    \end{aligned}
    \label{eq:derisotr2coinc}
\end{equation}
where $\mathcal{I}$ is the fourth-order identity tensor.
Finally, when all the eigenvalues coincide, Eq. \eqref{eq:isotr2coinc} reduces to:
\begin{equation}
    \tens{S}= \frac{I_{1S}}{3} \tens{I}
    \label{eq:isotr_3coinc}
\end{equation}
whilst it derivative can be computed by particularizing Eq. \eqref{eq:derisotr2coinc}.
By observing that when $\tens{t} \rightarrow \tens{0}$, $\tens{N}_i \rightarrow \tens{I}/3$, so that its deviatoric part $\hat{\tens{N}}^d\rightarrow \tens{0}$.
Observing that  $q_S \rightarrow 0$ when $q_T \rightarrow 0$, using a Tayor expansion for $q_T \rightarrow 0$, it will result:
\begin{equation}
    q_S(I_{1T}, 0)\approx \der{q_S}{q_T} q_T
    \nonumber
\end{equation}
so that $q_S/q_T \rightarrow \der{q_S}{q_T}$, and finally:
\begin{equation}
    \dert{\tens{S}}{\tens{\myT}}= \frac{1}{3}\der{I_{1S}}{I_{1T}} \tens{I} \otimes \tens{I}+ \der{q_S}{q_T} \left(\mathcal{I} - \frac{1}{3}   \tens{I} \otimes \tens{I}  \right)
    \label{eq:jacobian_3coinc}
\end{equation}

\section{Applications}
\subsection{Isotropic elastoplastic materials under small-strains and displacements}
Let consider a generic elastoplastic isotropic material, in which the principal directions of 
the elastic strains and of the stress coincides. 
Let be $\tens{s}$ the deviatoric part of the Cauchy stress tensor $\tens{\sigma}$, and
\begin{equation}
       p=\frac{1}{3}\tr{\tens{\sigma}}
       \nonumber
\end{equation}
\begin{equation}
       q=\sqrt{\frac{3}{2}\tens{s}:\tens{s}}
       \nonumber
       \end{equation}
\begin{equation}
       \theta_\sigma=\frac{1}{3} \arcsin\left( -\frac{27}{2} \frac{\det {\tens{s}}}{q^3}\right)
    \label{eq:stressinvar}
\end{equation}
the stress invariants, i.e.
the hydrostatic pressure, the equivalent von Mises stress, and the stress Lode's angle respectively.

In a general backward Euler integration scheme, let be $\tens{\eps}^*$ and $\Delta \tens{\eps}^p$ the elastic strain predictor and the plastic strain increment respectively. 
The plastic strain increment can be computed as a function of an \emph{isotropic} plastic potential $g(p,q,\theta_\sigma)$ as
\begin{equation}
\Delta \tens{\eps}^p= \der{g(p,q,\theta_\sigma)}{\tens{\sigma}}\Delta \gamma
\nonumber
\end{equation}
where $\Delta \gamma$ is the plastic multiplier. Since $g(p,q,\theta_\sigma)$ is an isotropic function of $\tens{\sigma}$, its derivative respect to $\tens{\sigma}$ will be co-axial with the stress \cite{bibbia} \cite{Panteghini2018}.
Then, since the elastic strain $\tens{\eps}^e$ is  co-axial with $\tens{\sigma}$ for the assumption of isotropy, it results that also
\begin{equation}
\tens{\eps}^*=\tens{\eps}^e+\Delta \tens{\eps}^p
\nonumber
\end{equation}
is co-axial with $\tens{\sigma}$. For these reasons, the principal directions of stress are \emph{a priori} known, being coincident with those of the predictor $\tens{\eps}^*$.
Let $\tens{e}^*$ the deviatoric part of the elastic predictor $\tens{\eps}^*$, and
\begin{equation}
\begin{gathered}
\eps_v^*=\tr{\tens{\eps}^*}\\
\eps_q^*= \sqrt{\frac{2}{3}\tens{e}^*:\tens{e}^* }\\
\theta^*_\eps=\frac{1}{3} \arcsin\left( - 4 \frac{\det {\tens{e^*}}}{\eps_q^{*3}}\right)
\end{gathered}
\nonumber
\end{equation}
the its invariants, i.e. the volumetric strain predictor, the equivalent von Mises strain predictor, and the strain predictor Lode's angle.

In general, if a standard return algorithm in the full tensorial space is employed, numerical problems and convergence difficulties can arise when two or more eigenvalues coincide. 
Instead,  $p$,  $q$, $\theta_{\sigma}$ can be more easily computed formulating a return algorithm in the invariants strain space \cite{Panteghini2018}. Once $p$, $q$ and $\theta_{\sigma}$ have been obtained as a function of the strain invariants predictor, it is necessary to compute the stress tensor $\tens{\sigma}$.
If $\eps_q^*\ne 0$ and $|\theta^*_\eps| \ne \pi/6$, one can compute the stress tensor from its invariants and from the eigenbasis $\tens{N}_i^*$ of the elastic strain predictor $\tens{\eps}^*$  by resorting to the 
spectral theorem. It results
\begin{equation}
\begin{aligned}
\tens{\sigma}= \sum \left[p(\eps_v^*,\eps_q^*,\theta^*_\eps) 
+ \frac{2}{3} q(\eps_v^*,\eps_q^*,\theta^*_\eps) \sin \beta_i (\eps_v^*,\eps_q^*,\theta^*_\eps)\right] \tens{N}_i^*
\end{aligned}
\nonumber
\end{equation}
where
\begin{equation}
\begin{gathered}
    \beta_{\rm{I}} = \theta_\sigma (\eps_v^*,\eps_q^*,\theta^*_\eps) + \frac{2}{3} \pi\\
    \beta_{\rm{II}} = \theta_\sigma(\eps_v^*,\eps_q^*,\theta^*_\eps)  \\
    \beta_{\rm{III}} = \theta_\sigma (\eps_v^*,\eps_q^*,\theta^*_\eps) - \frac{2}{3} \pi\\
    \end{gathered}
    \nonumber
\end{equation}
and $\tens{N}_i^*$ is computed from Eq. \eqref{eq:eigenbasis_impl} as a function of the invariants of $\tens{\eps}^*$ and its principal components.
The consistent jacobian matrix\footnote{It should be noted that this \emph{general} approach has been recently adopted by the Author in \cite{Panteghini2022}, while in his older work \cite{Panteghini2018}, in order to avoid the computation of the spin of the eigenbasis, the spectral representation of the stress was computed as a function of the eigenvectors of the strain predictor, while jacobian matrix was obtained by means of a "simplified" procedure based on the inversion of a $6$x$6$ matrix. Unfortunately, this procedure is model-specific and requires the smoothness in the deviatoric plane of the yield function and of the plastic potential.} can be computed from Eq. \eqref{eq:jacobian:compl} as
\begin{equation}
\begin{aligned}
\dert{\tens{\sigma}}{\tens{\eps}^*}=\sum \left[p+ \frac{2}{3} q \sin \beta_i \right] 
\dert{\tens{N}_i^*}{\tens{\eps}^*}
+
\tens{N}_i^* 
\\
\otimes
\left\{
\left[
\der{p}{\eps_v^*} 
+\frac{2}{3} \left(\der{q}{\eps_v^*} \sin \beta_i +
q \der{\theta_\sigma}{\eps_v^*} \cos \beta_i 
\right)
\right]\tens{I}
\right.
\\
\left.
+\frac{2}{3\eps_q^*}
\left[\der{p}{\eps_q^*}
+\frac{2}{3} \left(\der{q}{\eps_q^*} \sin \beta_i +
q  \der{\theta_\sigma}{\eps_q^*} \cos \beta_i
\right) \right]\tens{e}^* 
\right.\\
\left.
+\left[\der{p}{\theta^*_\eps}
+\frac{2}{3} \left(\der{q}{\theta^*_\eps} \sin \beta_i +
q  \der{\theta_\sigma}{\theta^*_\eps} \cos \beta_i
\right)  \right]\der{\theta^*_\eps}{\tens{\eps}^*}
\right\}
\end{aligned}
\nonumber
\end{equation}
where the eigenbasis spin $\dert{\tens{N}_i^*}{\tens{\eps}^*}$  and $\der{\theta^*_\eps}{\tens{\eps}^*}$  are computed as a function of the invariants and principal components of $\tens{\eps}^*$ from Eqs. \eqref{eq:spin} and  \eqref{eq:dertheta_impl} respectively.

If $\eps_q^*$ is not nil, at least two eigenvalues of the strain predictor $\tens{\eps}^*$ are distinct.
Specifically, if $\theta^*_\eps=\pm \pi/6$ two eigenvalues of $\tens{\eps}^*$ will be coincident. In this case, from Eq. \eqref{eq:spectral_invariants3} it will result that $\tens{e}^*$ will be \emph{proportional} to the deviatoric part of the eigenbasis associated to its non-repeated eigenvalue.  Hence, from Eq. \eqref{eq:isotr2coinc} one simply obtains:
\begin{equation}
\tens{\sigma}=p (\eps_v^*,\eps_q^*,\theta^*_\eps) \tens{I}+\frac{2 }{3  \eps_q^*} q (\eps_v^*,\eps_q^*,\theta^*_\eps) \tens{e}^*
\nonumber
\end{equation}
Also the eigenbasis of the deviatoric part of the plastic strain increment $\Delta \tens{e}^p$ and of the elastic strains will coincide with those of $\tens{e}^*$, and then it will result:
\begin{equation}
\Delta \tens{\eps}^p=\frac{\eps^p_v}{3} \tens{I}+\frac{\eps_q^p}{\eps_q^*}  \tens{e}^*
\nonumber
\end{equation}
\begin{equation}
\tens{\eps}^e=\frac{\eps^e_v}{3}+\frac{\eps_q^e}{\eps_q^*}  \tens{e}^*
\nonumber
\end{equation}
The jacobian matrix can be obtained simplifying Eq.  \eqref{eq:derisotr2coinc} using Eq.  \eqref{eq:spectral_invariants3}. It yields:
\begin{equation}
\begin{aligned}
\dert{\tens{\sigma}}{\tens{\eps}^*}=
\der{p}{\eps_v^*} \tens{I}\otimes \tens{I}+ \frac{2}{3\eps_q^*}
\left[\der{p}{\eps_q^*} \left(\tens{I}\otimes \tens{e}^*\right) 
+\der{q}{\eps_v^*} \left(\tens{e}^* \otimes \tens{I} \right) 
\right. \\ \left.
+ \frac{2}{3 \eps_q^*} \left(\der{q}{\eps_q^*} -\frac{q}{\eps_q^*} \right) \left( \tens{e}^* \otimes \tens{e}^*\right) 
+ q \left( \mathcal{I}-\frac{1}{3} \tens{I}\otimes \tens{I}\right)
\right]
\end{aligned}
\nonumber
\end{equation}

If $\eps_q^*$ is nil, the strain predictor $\tens{\eps}^*$ will be a volumetric tensor, since its spectral decomposition has the same structure of Eq. \eqref{eq:spectral3}. Moreover, $\eps_q^*=0$ implies $\tens{e}^* = \tens{0}$.
Since the material is isotropic, the eigenbasis of $\tens{\sigma}$ and $\tens{\eps}^*$ the same, resulting to be coincident with the second-order identity tensor $\tens{I}$. Then, from Eq. \eqref{eq:spectral3} it will result
\begin{equation}
\tens{\sigma}= p (\eps_v^*) \tens{I}
\nonumber
\end{equation}

The derivative of the eigenbasis is undefined. However, as explained in the section above, the Jacobian Matrix can be obtained as a limit case of Eq. \eqref{eq:derisotr2coinc}, i.e., using Eq. \eqref{eq:jacobian_3coinc}.
Let observe that, under purely volumetric conditions, the convexity of the elastic potential requires \cite{Lagioia2019}:
\begin{equation}
    \der{p}{\eps_q^*} =\der{q}{\eps_v^*}=0
\nonumber
\end{equation}
It results:
\begin{equation}
\begin{aligned}
\dert{\tens{\sigma}}{\tens{\eps}^*}=
\der{p}{\eps_v^*} \tens{I}\otimes \tens{I}+ \frac{2}{3}
\der{q}{\eps_q^*}\left( \mathcal{I}-\frac{1}{3} \tens{I}\otimes \tens{I}\right)
\end{aligned}
\nonumber
\end{equation}

\subsection{Computation of logarithmic strain tensor from displacement gradient}

In the framework of large strains and rotations, let $\bvec{p}$ denotes the reference coordinate system.
Indicating with $\bvec u(\bvec p)$ the vector function describing the displacement of each material point, it results that its final position will be (i.g. \cite{bibbia})
\begin{equation}
    \bvec{x}=\bvec{p}+\bvec u(\bvec p)
    \nonumber
\end{equation}
The deformation gradient $\tens F$ is defined as
\begin{equation}
    \tens{F}= \nabla_p \bvec{x} = \tens{I}+\nabla_p \bvec u(\bvec p)
    \nonumber
\end{equation}
By applying the \emph{polar decomposition} (i.g. \cite{bibbia}) to the deformation gradient $\tens{F}$, one obtains:
\begin{equation}
    \tens{F} = \tens{V}\tens{R} 
    \nonumber
\end{equation}
where the orthogonal tensor $\tens{R}$ describes the local rotation, whilst the symmetric positive definite tensor
$\tens{V}$ is the left stretch tensor, where
\begin{equation}
     \tens{V}^2 = \tens{B} = \tens{F} \tens{F}^T
    \nonumber
\end{equation}
$\tens{B}$ being the left Cauchy-Green tensor.
The logarithmic strain tensor can be computed as:
\begin{equation}
    \tens{\eps}=\ln \tens{V}=\frac{1}{2} \ln \tens{B}
\nonumber
\end{equation}
i.e., 
\begin{equation}
    \tens{\eps}= \frac{1}{2} \sum \ln\left( \lambda^B_i\right) \tens{N}^B_{i}
    \label{eq:spectraleps}
\end{equation}
where $\lambda^B_i$ and $\tens{N}^B_{i}$ are the $i-$th principal component and eigenbasis of the tensor $\tens{B}$ respectively.

The invariants of $\tens{B}$, $I_{1B}$, $J_{2B}$
and $\theta_B$ can be computed using Eqs. \eqref{eq:I1}, \eqref{eq:defj2} and \eqref{eq:deftheta}, whilst the principal components $\lambda^B_i$ can be obtained using Eqs. \eqref{eq:lambda}.

If $\lambda^B_i$ are distinct, i.e., if $J_{2B} \ne 0$ and $|\theta_B| \ne \pi/6$, all the eigenbasis $\tens{N}^B_{i}$ of the left Cauchy-Green tensor can be computed as a function of its invariants and its principal components using Eq. \eqref{eq:eigenbasis_impl}. The logarithmic strain tensor can be computed using Eq. \eqref{eq:spectraleps}. 
The jacobian matrix $\dert{\tens{\eps}}{\tens{B}}$ can be computed by using Eq. \eqref{eq:jacobian:compl}:
\begin{equation}
    \dert{\tens{\eps}}{\tens{B}}=\frac{1}{2} \sum \left[ \ln\left( \lambda^B_i\right) 
    \dert{\tens{N}^B_{i}}{\tens{B}}
       + \frac{1}{\lambda^B_i} \tens{N}^B_{i}\otimes \tens{N}^B_{i} \right]
       \nonumber
\end{equation}
where $\dert{\tens{N}^B_{i}}{\tens{B}}$ can be computed using Eq. \eqref{eq:spin}.

When two principal components of $\tens{B}$ are coincident, i.e. if $J_{2B} \ne 0$ and $|\theta_B| = \pi/6$, one can compute $\tens{\eps}$ by exploiting the proportionality between the deviatoric part $\tens{b}$ of $\tens{B}$ and $\tens{e}$.
Let start by computing the invariants $q_{\eps}=\sqrt{3 J_{2\eps}}$ and $I_{1\eps}$ of $\tens{\eps}$ as a function of $q_B= \sqrt{3 J_{2B}}$ and $I_{1B}$. Let observe that it results
\begin{equation}
    q_B = \pm \left( \lambda^B_{\rm{II}}-\hat \lambda^B\right) \;\; \mbox { for }\;\; \theta_B = \pm \frac{\pi}{6}
    \nonumber
\end{equation}
By solving this expression for $\lambda^B_{\rm{II}}$ one obtatins
\begin{equation}
    \lambda^B_{\rm{II}}= \hat \lambda^B \pm q_B \;\; \mbox { for }\;\; \theta_B = \pm \frac{\pi}{6}
    \label{eq:lambdaB2temp}
\end{equation}
Substituting this result into the definition of $I_{1B}=\hat \lambda^B + 2\lambda^B_{\rm{II}}$ and solving for $\hat \lambda^B$ gives
\begin{equation}
     \hat \lambda^B= \frac{ I_{1B}\mp 2 q_B}{3}  \;\; \mbox { for }\;\; \theta_B = \pm \frac{\pi}{6}
     \nonumber
\end{equation}
By substituting this expression into Eq.  \eqref{eq:lambdaB2temp} one obtains
\begin{equation}
    \lambda^B_{\rm{II}}=  \frac{ I_{1B}\pm  q_B}{3}  \;\; \mbox { for }\;\; \theta_B = \pm \frac{\pi}{6}
    \nonumber
\end{equation}
One can now compute the invariants of $\tens{\eps}$ as a function of those of $\tens{B}$. It results:
\begin{equation}
    \begin{aligned}
    \begin{aligned}
    I_{1\eps} =\hat \lambda^\eps+ 2 \lambda^\eps_{\rm{II}} = \frac{1}{2} \left[ 
        \ln \left( \frac{ I_{1B}\mp 2 q_B}{3}\right)+2\ln \left(  \frac{ I_{1B}\pm  q_B}{3}\right)
        \right],
    \\
        q_{\eps}= \pm \left( \lambda^\eps_{\rm{II}}-\hat \lambda^\eps\right) = \pm \frac{1}{2} \left( \ln \lambda^B_{\rm{II}} - \ln \hat \lambda^B \right) = \pm \frac{1}{2} \ln \left( \frac{I_{1B}\pm  q_B}{I_{1B}\mp 2 q_B}\right)\\
        \end{aligned}
        \\
     \;\; \mbox { for }\;\;  \theta_\eps = \theta_B = \pm \frac{\pi}{6}
     \end{aligned}
     \label{eq:invEinvB}
\end{equation}
The logarithmic strain tensor $\tens{\eps}$ can be finally computed using Eq.\eqref{eq:isotr2coinc}. It results:
\begin{equation}
    \tens{\eps} = \frac{I_{1\eps}}{3} \tens{I} + \frac{q_\eps}{q_B} \tens b
    \nonumber
\end{equation}
Its derivative can be obtained by applying Eq. \eqref{eq:derisotr2coinc}. It results:
\begin{equation}
\begin{aligned}
    \dert{\tens{\eps}}{\tens{B}}= \frac{1}{3}\der{I_{1\eps}}{I_{1B}} \tens{I} \otimes \tens{I} \mp \frac{1}{2} \der{I_{1\eps}}{q_B} \tens{I} \otimes \hat{\tens{N}}^d_B
    + \frac{q_\eps}{q_B}\left(\mathcal{I} - \frac{1}{3}   \tens{I} \otimes \tens{I}  \right)\\
    + \frac{3}{2} \left(\der{q_\eps}{q_B} - \frac{q_\eps}{q_B} \right) \hat{\tens{N}}^d_B \otimes \hat{\tens{N}}^d_B
        \mp\der{q_\eps}{I_{1B}}  \hat{\tens{N}}^d_B \otimes \tens{I}\;\; \mbox { for }\;\;\theta_\eps = \theta_B = \pm \frac{\pi}{6}
    \end{aligned}
    \nonumber
\end{equation}    
where, from Eq. \eqref{eq:spectral_invariants3}:
\begin{equation}
    \hat{\tens{N}}^d_B= \mp \frac{1}{q_B} \tens{b} \;\; \mbox { for }\;\; \theta_B = \pm \frac{\pi}{6}
\nonumber
\end{equation}
and, by computing the derivatives of Eq. \eqref{eq:invEinvB}:
\begin{equation}
\begin{aligned}
     \begin{aligned}
     &\der{I_{1\eps}}{I_{1B}}=\frac{3(\pm  q_B - I_{1B})}{(I_{1B}\pm q_B)(\pm 4 q_B -2 I_{1B})}\\
     &\der{I_{1\eps}}{q_B}=\frac{3 q_B}{(\pm 2 q_B-I_{1B})(I_{1B}\pm q_B)}\\
     &\der{q_\eps}{I_{1B}} = \frac{3 q_B}{(I_{1B} \pm q_B)(\pm 4 q_B -2 I_{1B})}\\
     &\der{q_\eps}{q_B} = - \frac{3 I_{1B}}{(I_{1B} \pm q_B)(\pm  q_B -2 I_{1B}))}\\   
     \end{aligned}
      \mbox { for } \theta_\eps = \theta_B = \pm \frac{\pi}{6}\\
      \end{aligned}
     \label{eq:derInvEpsInvB}
\end{equation}
Finally, if $J_{2B} = 0$, then the logarithmic strain will be purely volumetric, and it will result $\lambda_i^B = \lambda^B$.
Eqs. \eqref{eq:invEinvB} become:
\begin{equation}
    \begin{aligned}
    & I_{1\eps} = \frac{3}{2} \ln \left( \frac{ I_{1B}}{3}\right) = \frac{3}{2} \ln \lambda^B
    \\
        &q_{\eps}= 0\\
        \end{aligned}
        \label{eq:InvEps_3Coinv}
\end{equation}
By applying Eq. \eqref{eq:isotr_3coinc} it will result:
\begin{equation}
\tens{\eps}=
\frac{1}{2}\ln \lambda^B \tens{I}
\nonumber
\end{equation}
To compute the derivative of $\tens{\eps}$ with respect to $\tens{B}$, let start substituting Eqs. \eqref{eq:InvEps_3Coinv} into Eqs. \eqref{eq:derInvEpsInvB}. It results:
\begin{equation}
  \begin{aligned}
     &\der{I_{1\eps}}{I_{1B}}=\frac{3}{2 I_{1B}} = \frac{1}{2 \lambda^B}\\
     &\der{I_{1\eps}}{q_B}=\der{q_\eps}{I_{1B}} = 0 \\
     &\der{q_\eps}{q_B} = \frac{3}{2 I_{1B}} = \frac{1}{2 \lambda^B}
     \end{aligned}   
     \nonumber
\end{equation}
By substituting these expressions into Eq. \eqref{eq:jacobian_3coinc} one obtains:
\begin{equation}
    \dert{\tens{\eps}}{\tens{B}}= \frac{1}{2\lambda^B} \mathcal{I}
    \nonumber
\end{equation}

\section{Conclusions}
The spectral representation of a symmetric, second-order tensor is an important tool in 
many applications of computational mechanics. 

While the computation of the eigenvalues of a symmetric, second-order tensor is a relative simple task, obtaining a closed-form expression for the eigenbasis is more complicate, especially when some eigenvalue is repeated. Moreover, in many computational mechanics applications, also the derivative of the spectral representation is required. The exact closed-form expressions available in the literature for both the eigenbasis and their derivative are quite hard to implement (see, e.g., \cite{bibbia}). For this reason, many Authors suggest to resort to series expansions, that however are available only specific functions (see, e.g., \cite{deSouzaNeto2001}, \cite{Ortiz2001}) or 
require automatic differentiation techniques for a generic function \cite{HUDOBIVNIK2016}, 

These approximate techniques are hard to apply 
when the isotropic tensor-valued functions are not known explicitly, such as, for instance, 
in the numerical integration of elastoplastic isotropic constitutive laws formulated in invariants space (\cite{Borja2003} \cite{Panteghini2018} \cite{Panteghini2022}).

In this paper, starting from a incidental result reported by Ogden \cite{Odgen1984} working only in the case of not coincident eigenvalues, an exact, simple and clear approach has been developed. Differently from that described by Miehe \cite{Miehe1993}, \cite{Miehe1998}  no particular requirements about the invertibility of the tensor, or its eigenvalues multiplicity  are necessary.

Two applications have been presented: (i) the  computation of stress tensor and of the stiffness matrix in the case of the numerical integration of an elastoplastic isotropic material in the invariant stress space, and (ii) the calculation of the logarithmic strain tensor from the displacement gradient, as well as its derivative with respect to the left Cauchy-Green tensor.

\bibliographystyle{plain}

\bibliography{eigenbasis} 

\end{document}